\documentclass[12pt]{article}
\usepackage{scicite}
\usepackage{times}
\usepackage{graphicx}
\usepackage{subfigure}
\usepackage{color}
\usepackage{url}
\usepackage[normalem]{ulem}
\newenvironment{sciabstract}{
\begin{quote} \bf}
{\end{quote}}

\title{An extremely energetic cosmic ray observed by a surface detector array}

\author{
Telescope Array Collaboration$^{\ast\dagger}$ \\
\normalsize{$^\ast$ Corresponding author: Toshihiro Fujii (toshi@omu.ac.jp)}\\
\normalsize{$^\dagger$ Telescope Array Collaboration authors and affiliations are listed in the end of this article.}\\
\normalsize{Published in \textit{Science} \url{https://doi.org/10.1126/science.abo5095}}
}

\date{}

\begin{document}
\baselineskip24pt
\maketitle

\begin{sciabstract}
  Cosmic rays are energetic charged particles from extraterrestrial sources, with the highest energy events thought to come from extragalactic sources.
  Their arrival is infrequent, so detection requires instruments with large collecting areas.
  In this work, we report the detection of an extremely energetic particle recorded by the surface detector array of the Telescope Array experiment.
  We calculate the particle's energy as 244\,$\pm$\,29\,(stat.) $^{+51}_{-76}$\,(syst.)\,exa-electron volts ($\sim$ 40\,joules). 
  Its arrival direction points back to a void in the
  large-scale structure of the Universe.
  Possible explanations include a large deflection by the foreground magnetic field, an unidentified source in the local extragalactic neighborhood or 
  an incomplete knowledge of particle physics.
\end{sciabstract}

Ultra-high-energy cosmic rays (UHECRs) are subatomic particles from extragalactic sources with energies greater than $10^{18}$ eV (1 exa-electron volt, EeV),
 about a million times higher than reached by human-made accelerators. 
The origins of UHECRs are thought to be related to the most energetic phenomena
in the Universe, such as relativistic jets and outflows associated with black holes, gamma-ray bursts and relativistic flares of active galactic nuclei, or large-scale accretion shocks around clusters of galaxies~\cite{Hillas:1985is}. 
Alternatively, UHECRs might be produced by physics beyond the Standard Model of particle physics~\cite{Bhattacharjee:1999mup,Stecker:2009hj,Berezinsky:1997hy},
though this possibility is constrained by upper bounds on the flux of ultra-high-energy (UHE) photons~\cite{TelescopeArray:2018rbt, PierreAuger:2022uwd}.
The acceleration mechanisms of these particles are also unknown. 
Because cosmic rays are charged, they are deflected along their path to Earth by intervening Galactic and extragalactic magnetic fields, so their arrival directions do not necessarily point to their sources.

Although cosmic rays with energy $>$100\,EeV have been observed~\cite{Linsley:1963km}, interactions with the cosmic microwave background radiation (CMB)~\cite{Penzias:1965wn} are expected to suppress the flux of UHECRs above 60\,EeV~\cite{bib:gzk1,bib:gzk2}.
This is because interactions between UHECR protons and the CMB can produce pions or induce photo-disintegration of heavy nuclei.
The resulting break in the expected energy spectrum is known as the Greisen-Zatsepin-Kuzmin (GZK) cutoff~\cite{bib:gzk1,bib:gzk2}.
This cutoff limits the origins of the highest energy particles detected on Earth to sources within 50 to 100\,mega-parsecs (Mpc) since they have a short enough path length to survive passage through the intergalactic medium.
At these distance scales, the Universe is not homogeneous: matter is concentrated in a large-scale structure (LSS) composed of galaxy clusters, superclusters, filaments and sheets, separated by intergalactic voids.
A suppression of cosmic ray flux at the highest energies consistent with the GZK cutoff has been observed ~\cite{bib:hires_gzk,AbuZayyad:2012ru,PierreAuger:2020kuy}.
However, UHECRs with energies above the GZK cutoff are expected to be deflected less strongly by magnetic fields due to their high kinetic energies, so their arrival directions are expected to be more closely correlated with their sources.

\section*{The Telescope Array experiment}
At energies greater than 100\,EeV, the flux of cosmic rays is less than one
particle per century per square kilometer~\cite{AbuZayyad:2012ru}.
This low flux can only be measured by an instrument with a collecting area of $\sim$1000\,km$^2$. 
The energy, mass and arrival direction of UHECRs can be reconstructed from the cascades of secondary particles (Extensive Air Shower, EAS) produced by their interaction with Earth's atmosphere.
Arrays of detectors such as plastic scintillators or water-Cerenkov stations are conventionally used to measure EASs at ground.

The Telescope Array experiment (TA) is a cosmic ray detector in the northern hemisphere, located at 39.30$^{\circ}$ North, 112.91$^{\circ}$ West and 1370\,m above sea level in Utah, the United States of America.
It consists of a surface detector array (SD) with 507 stations arranged in a square grid. Each detector has two 3\,m$^2$ layers of plastic scintillator which detect charged EAS particles. The stations are spaced by 1.2\,km, giving a total effective area of 700\,km$^2$~\cite{AbuZayyad:2012kk}. 
The time-dependent response of the surface detectors is continuously monitored and calibrated by penetrating muons and electrons~\cite{AbuZayyad:2012kk}. 
The sky over the SD is viewed by fluorescence detectors which directly measure EAS developments in the atmosphere and provide a calorimetric measurement of the shower energy~\cite{Tokuno:2012mi}.
The mass of the primary cosmic ray is estimated using the fluorescence detectors by determining $X_{\max}$, the atmospheric slant depth (measured in g/cm$^2$) at which an EAS deposits most of its energy.
The $X_{\max}$ observable is related to the mass composition by a statistical analysis. 
TA does not determine the particle mass on an event-by-event basis~\cite{TelescopeArray:2018xyi}.
The SD measurements carry indirect information about the primary composition, 
which is extracted on a statistical basis using machine learning~\cite{TelescopeArray:2018bep}.

The arrival direction of a cosmic-ray particle is determined from the relative arrival times of the shower front at multiple SD stations,
measured by a time-synchronized global positioning system (GPS) module mounted on each station. 
The particle density measured at a distance of 800\,m from the EAS axis, $S_{800}$, is used as the energy indicator.
$S_{800}$ is converted to the primary energy of the cosmic ray as a function of zenith angle based on Monte Carlo EAS simulations~\cite{Heck:1998vt}. 
The SD energy scale was calibrated to the calorimetric energy measured by the fluorescence detectors with a factor of 1/1.27~\cite{AbuZayyad:2012ru}.
The resolution of the SD is 1.5$^{\circ}$ in arrival direction and 15\% in energy~\cite{AbuZayyad:2012ru}, with a systematic uncertainty of 21\%~\cite{TheTelescopeArray:2015mgw}.
The detailed analysis procedure and event reconstruction are described in ~\cite{sm}.

\begin{figure}
  \subfigure{\includegraphics[width=0.48\linewidth]{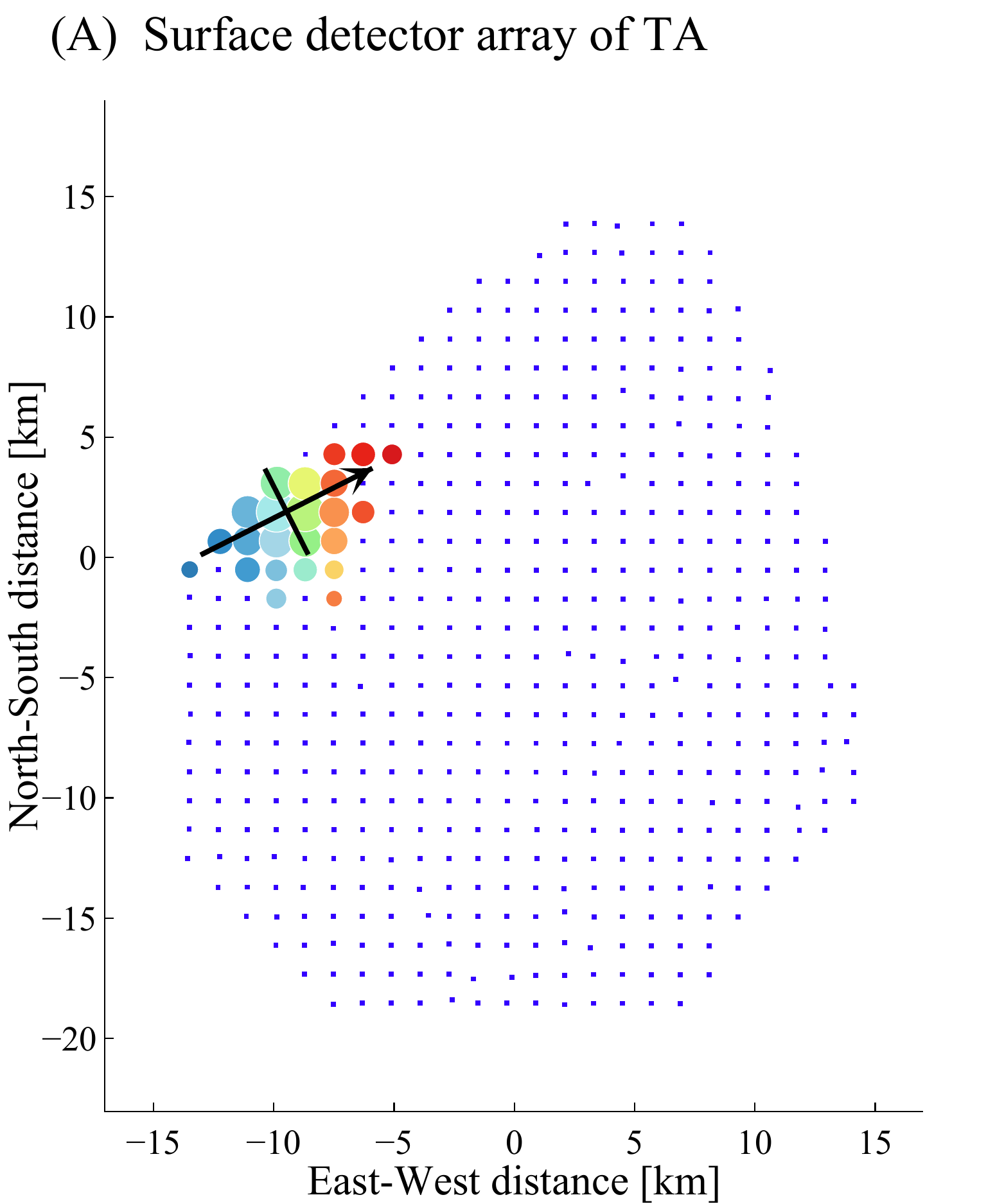}}
  \subfigure{\includegraphics[width=0.50\linewidth]{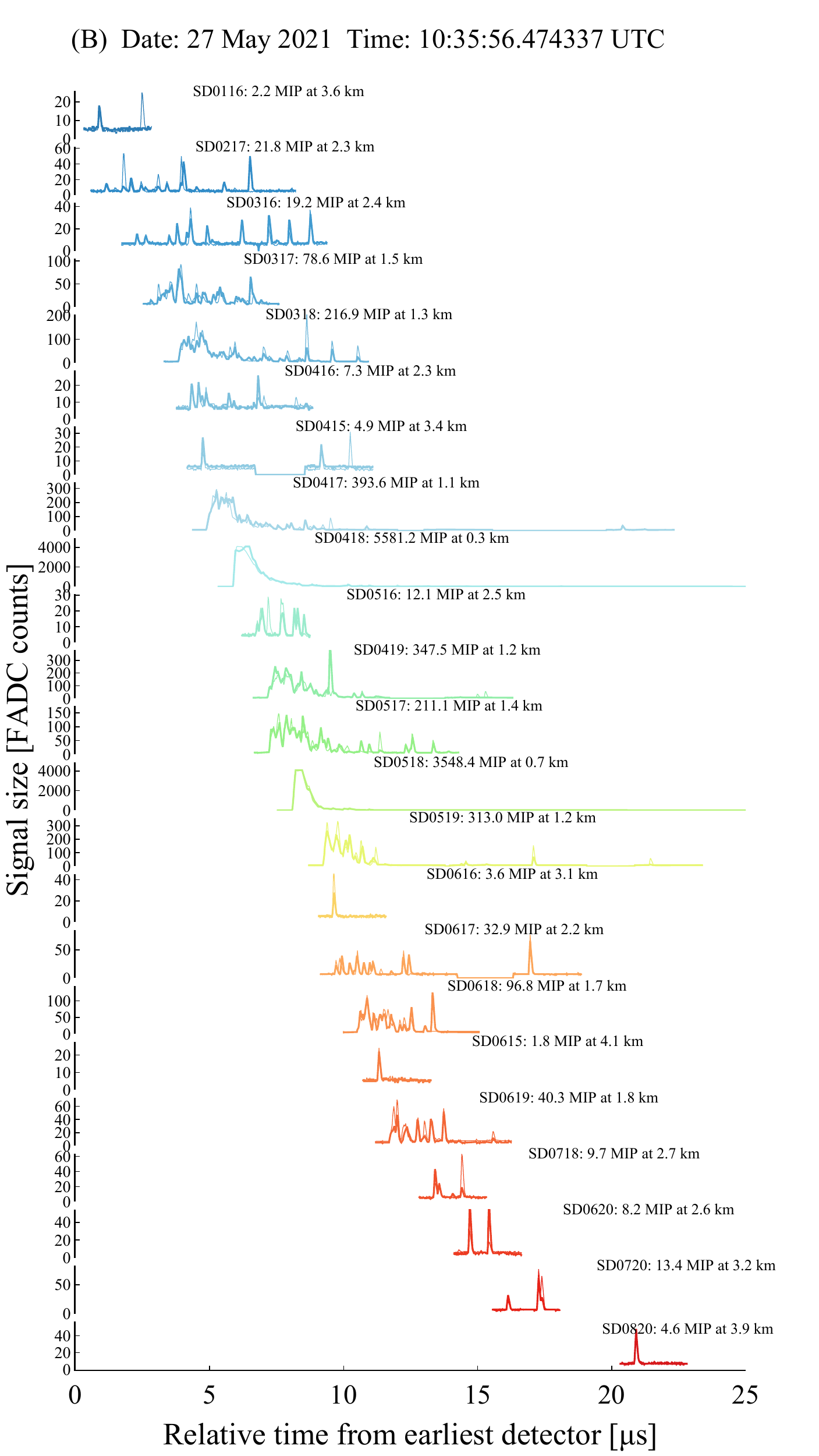}}
  \caption{
  \textbf{The high-energy particle event observed by TA SD on 27 May 2021.} 
  (A) A map of the TA SD. Each dot indicates the location of a SD station. 
  The black arrow indicates the shower direction projected on the ground. 
  The landing shower core position is located at ($-$9471 $\pm$ 31\,m, 1904 $\pm$ 23\,m) from the central position of the SD.
  The size of the colored circles is proportional to the number of particles detected by each station and the color denotes the relative time from the earliest detector.
  (B) The corresponding detector waveforms of flash analog-to-digital converter (FADC) counts for each station are shown with total signal in units of the minimum ionizing particle (MIP) and distance from the shower axis indicated. Thick and thin lines are recorded signals in upper and lower layers for each station. 
  Each SD is identified by a four digit number. The first two digits correspond to the column of the array in which the SD is located (numbered west to east) and the second two digits correspond to the row (numbered south to north). 
  The coordinated universal time (UTC) is used for the date and time of the event. The size of particle density and color code of the relative time in (A) are quantitatively indicated for each detector.
  }
  \label{fig:eventdisplay}
\end{figure}

\section*{Energetic particle on 27 May 2021}
An unusually high energy event was identified during an arrival direction analysis~\cite{bib:hotspot_ta} of all SD data taken between May 2008 and November 2021.
This event triggered 23 detectors at the north-west region of the TA SD. 
To determine $S_800$ and the primary energy, the lateral density distribution is shown in ~\cite{sm}. 
Applying the defined analysis procedure, the event on 27 May 2021 has a reconstructed energy of 244\,$\pm$\,29\,(stat.)\,$^{+51}_{-76}$\,(syst.)\,EeV in the detector frame. 
This energy is $\sim4 \times 10^{7}$ times higher than the $\sim$7\,tera-electron volts (TeV) protons accelerated by the Large Hadron Collider (LHC)~\cite{LyndonEvans_2008}. 
When this cosmic ray particle experienced its first collision with a nucleon at rest in the upper atmosphere, the corresponding center-of-mass energy of the particle collision, assuming the particle was a proton, reached $\sim$700\,TeV. 

\begin{table}
  \centering
  \caption{\textbf{The reconstructed properties of the high-energy event.} The energy and the particle density at the distance of 800\,m, $S_{800}$, of the high-energy particle. Also listed are its arrival direction in the zenith-azimuth coordinates and the derived equatorial coordinates. The azimuth angle is defined
  to be anti-clockwise from the East.}
  \begin{tabular}{c|c|c|c|c|c|c}
    Date & Energy & $S_{800}$ & Zenith & Azimuth & Right & Declination \\
    (UTC) & (EeV) & (m$^{-2}$) & angle & angle & Ascension & \\ \hline \hline
    27 May 2021  & 244$\pm$29(stat.) & 530$\pm$57 & 38.6$\pm$0.4$^{\circ}$  & 206.8$\pm$0.6$^{\circ}$  & 255.9$\pm$0.6$^{\circ}$  & 16.1$\pm$0.5$^{\circ}$ \\
    10:35:56 & $^{+51}_{-76}$\,(syst.) & & & & & \\ \hline
  \end{tabular}
  \label{tab:highest}
\end{table}

Fig.~\ref{fig:eventdisplay}A shows a map of the TA SD and its recorded signals for the high-energy cosmic ray, including the footprint of the EAS on the TA SD.
Fig.~\ref{fig:eventdisplay}B shows the recorded signal size measured at each surface detector. 
Table~\ref{tab:highest} summarizes the reconstructed properties of the event. 
The waveforms recorded by detectors at distances above 2\,km contain many peaks from muons induced by the hadronic interactions. 
With that many muon components, the primary particle is unlikely to be a photon because EASs induced by photons primarily consist of electromagnetic particles.
We applied a neural network proton-photon classifier developed for photo-induced shower searches using the TA SD ~\cite{TAgamma_ICRC2021,TAML_ICRC2021} to this event. 
We find the classifier excludes a photon as the primary particle at the 99.986\% confidence level, instead favoring a primary proton.
However, the classifier cannot distinguish between protons and heavier nuclei because the fluorescence detectors were not operating during this event due to bright moon light.

The core position of this event was located 1.1\,km from the north-west edge of the SD (Fig.~\ref{fig:eventdisplay}A). 
We evaluate the statistical uncertainty of the reconstructed energy using a detector simulation~\cite{AbuZayyad:2012ru} assuming the reconstructed geometry and energy parameters; we find an energy resolution of 29\,EeV for this event.
Assuming an energy spectrum of $E^{-4.8}$ above 100\,EeV as measured by the TA SD~\cite{AbuZayyad:2012ru}, 
a migration effect where lower energy showers are reconstructed with higher energies due to the energy resolution is evaluated as $-3$\%.
We include an additional systematic uncertainty due to unknown primary of $-10$\% in the direction of lower energies as calculated from simulations~\cite{sm}.
There was no lightning or thunderstorm activity recorded in the vicinity of the TA site on 27 May 2021~\cite{vaisala}.

\section*{Comparison with previous events}
Previously reported extremely high-energy cosmic ray events include a 320\,EeV particle in 1991~\cite{Bird:1994uy}, a 213\,EeV particle in 1993~\cite{Hayashida:1994hb} and a 280\,EeV particle in 2001~\cite{Sakaki:2001md}. The 1991 event was measured using fluorescence detectors whilst the 1993 and 2001 events were both detected using surface detector arrays. All of these events were recorded by detectors in the northern hemisphere. 
A search in the southern hemisphere has not identified any events with energy greater than 166 EeV~\cite{PierreAuger:2022axr}, although there is an energy scale difference between the experiments~\cite{spectrumwg_ptep}.
Although the event we have detected was measured with a surface detector array, the reported energy of 244\,EeV has been normalized to the equivalent energy that would have been measured with the TA fluorescence detector and is thus comparable to the 1991 event. This normalization is done as fluorescence detectors provide a direct, calorimetric measurement of the shower energy. The original TA SD reconstructed energy of 309$\pm$37(stat.)\,EeV~\cite{sm} is comparable to the 1993 and 2001 events.


\section*{Possible sources of the cosmic ray}

\begin{figure}
 \includegraphics[width=1\linewidth]{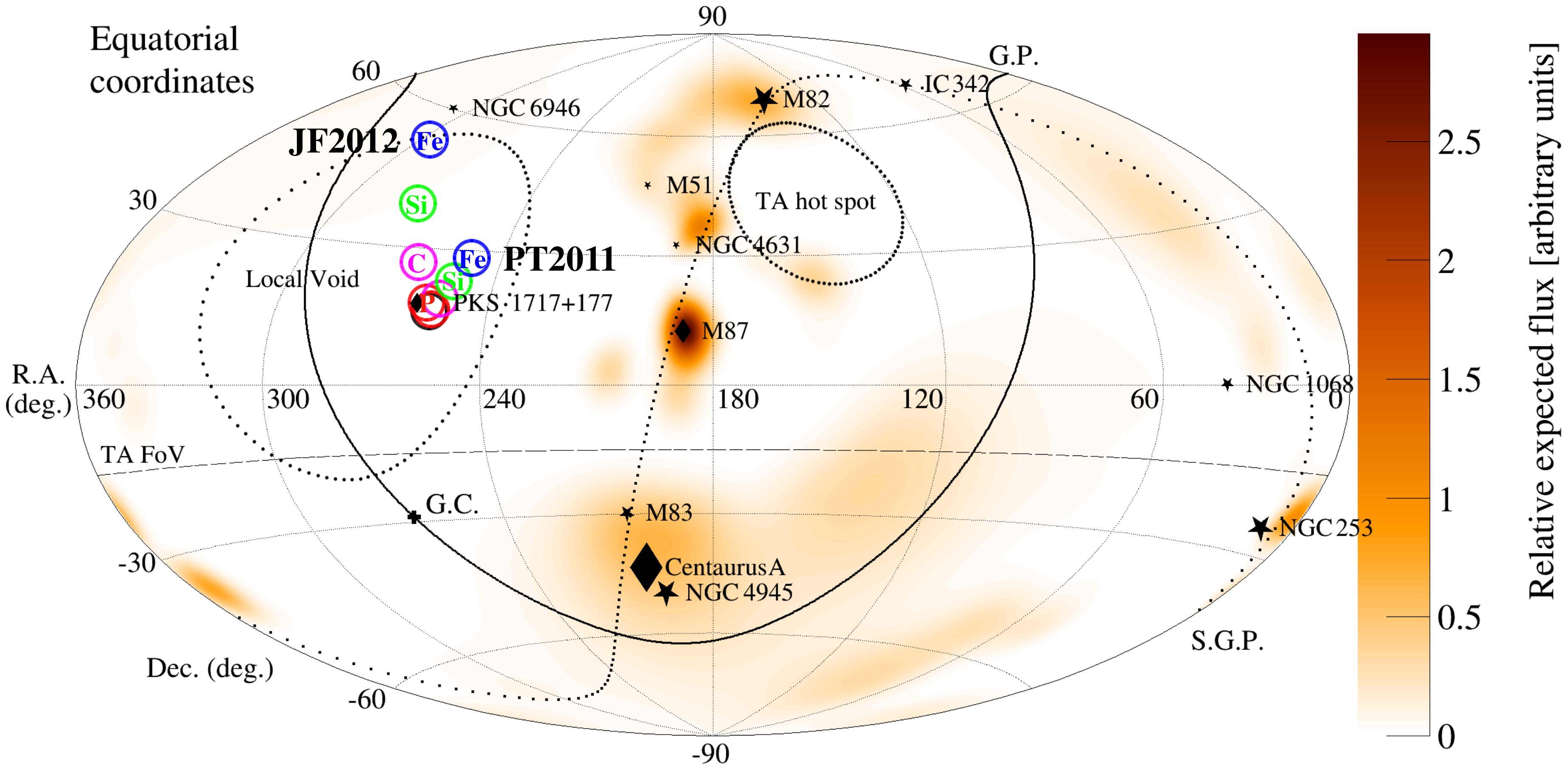}
 \caption{\textbf{Sky map in equatorial coordinates showing the arrival direction of the 27 May 2021 high-energy cosmic ray particle (black circle).} Also shown are calculated back-tracked directions assuming two models of the Milky Way regular magnetic field labelled JF2012~\cite{Jansson:2012pc} and PT2011~\cite{Pshirkov:2011um}.
 For each model, different symbols indicate the directions calculated for four possible primary species: proton (P, red), carbon (C, purple), silicon (Si, green), and iron (Fe, blue).
 The color bar indicates the relative flux expected from the inhomogeneous source-density distribution in the local LSS, smeared with a random Milky Way magnetic field. 
 For comparison, nearby gamma-ray-emitting active galactic nuclei are shown by
 filled diamonds, and nearby starburst galaxies are marked with filled stars, both with sizes that scale by the expected flux shown in ~\cite{bib:sbg_auger}. 
 The closest object to the proton back-tracked direction in a gamma-ray source  catalog~\cite{Fermi-LAT:2019yla} is the active galaxy PKS 1717+177. 
 The dotted large circle centered around (R.A., Dec.) = (146.7$^{\circ}$, 43.2$^{\circ}$) indicates the previously reported TA hot spot~\cite{bib:hotspot_ta} and the dashed horizontal curve indicates the limit of the TA field-of-view (FoV). The left dotted large circle centered around around (R.A., Dec.) = (279.5$^{\circ}$, 18.0$^{\circ}$) is a location of the Local Void~\cite{Tully:2007ue}.
 The galactic plane (G.P.) and the super-galactic plane (S.G.P.) are shown as solid and dotted curves, respectively. The galactic center (G.C.) is indicated by the cross symbol.}
\label{fig:highest}
\end{figure}

Fig.~\ref{fig:highest} shows the calculated arrival direction of the 27 May 2021 event in equatorial coordinates.
The arrival direction is not far from the disk of the Milky Way, where the
Galactic magnetic field (GMF) is strong enough to substantially deflect
even a particle with an energy of 244~EeV, especially if the primary particle is a heavy nucleus with a large electric charge. 
The map also shows the back-tracked arrival direction calculated assuming two GMF models~\cite{Jansson:2012pc,Pshirkov:2011um} and four possible primary particles (proton, carbon nucleus, silicon nucleus or iron nucleus). 
We used the back-tracking method of a cosmic-ray propagation framework~\cite{AlvesBatista:2016vpy} to find the arrival direction for the cosmic ray before it entered the Milky Way. 

We compared the arrival directions to a catalog of gamma-ray sources~\cite{Fermi-LAT:2019yla}. We find the active galaxy PKS 1717+177 is located within 2.5$^{\circ}$
of the calculated direction for a proton primary. 
PKS\,1717+177 is a flaring source~\protect\cite{Fermi-LAT:2019yla}, which has been proposed as potential cosmic ray sources~\cite{Farrar:2008ex}.
However, its distance of $\sim$ 600\,Mpc (corresponding to a redshift of 0.137)~\cite{Sowards-Emmerd_2005} is expected to be too large for UHECR propagation to Earth, because the average propagation distance at an energy of 244\,EeV is calculated to be $\sim30$\,Mpc for both proton and iron primaries~\cite{sm}. 
We therefore disfavor PKS\,17171+177 as the source of this event.

Fig.~\ref{fig:highest} also shows the relative expected flux from an inhomogeneous source-density distribution following the local LSS~\cite{MapConstruction}, weighted by the expected attenuation for a 244\,EeV iron primary smoothed to reflect the smearing due to turbulent magnetic fields in the Milky Way~\cite{sm}.
Also shown are nearby gamma-ray emitting active galactic nuclei and starburst galaxies, which have been proposed as possible cosmic ray sources~\cite{bib:sbg_auger, Abbasi:2018tqo}.
The arrival direction of this event is consistent with the location of the Local Void, a cavity between the Local Group of galaxies and nearby LSS filaments~\cite{Tully:2007ue}. 
There are only a small number of known galaxies in the void, none of which are expected sites of UHECR acceleration. 
Even taking into account the range of possible GMF deflections and primary mass, we do not identify any candidate sources for this event.
Only in the JF2012 GMF model assuming an iron primary does the source direction approach a part of the LSS populated by galaxies.
The starburst galaxy NGC\,6946, called the Fireworks Galaxy, at the distance of 7.7\,Mpc~\cite{10.1093/mnrasl/slz030} is close to this backtracked direction.
However, NGC\,6946 is not detected in gamma-rays, so is unlikely to be a strong source of UHECRs. 

If the energy of this event was close to the lower bound of its uncertainties, then the average propagation distance is longer than we assumed in Fig. 2 and the deflection in the GMF would be larger (Fig. S3).
This effect would increase the number of possible source galaxies, assuming a steady source (Supplementary Text).
For the alternative case of transient sources, identifying a source is complicated by the time delays between electromagnetic radiation and charged particles, due to the additional path lengths induced by magnetic deflection. 
We therefore cannot identify any potentially related transient sources. 

Nevertheless, the detection of this highly energetic particle  
allows us to estimate $D_0$, the distance to the closest UHECR source.
Assuming that the particle is an iron nucleus injected with an initial energy $E_0 = 10^3$\,EeV, taking into account the energy loss length estimated by the same propagation framework used in the back-tracking method~\cite{Kalashev:2014xna}, we find $D_0 = 10.3^{+5.3}_{-3.0}$\,Mpc.
Alternatively, assuming a proton primary, we find $D_0 = 27.0^{+3.8}_{-3.0}$\,Mpc (Supplementary Text).
At these energies, the UHECR background of distant sources is attenuated by the energy loss length, so only sources from the local Universe can contribute. 
We set upper limits on the deflection by assuming a maximum value of the turbulent extragalactic magnetic field $B_{\rm rms}\sim1$\,nG and a 1\,Mpc characteristic length scale, finding $<$20$^\circ$ for iron and $<$1$^\circ$ for proton.

\section*{Distribution of other TA events}
\begin{figure}
  \includegraphics[width=1\linewidth]{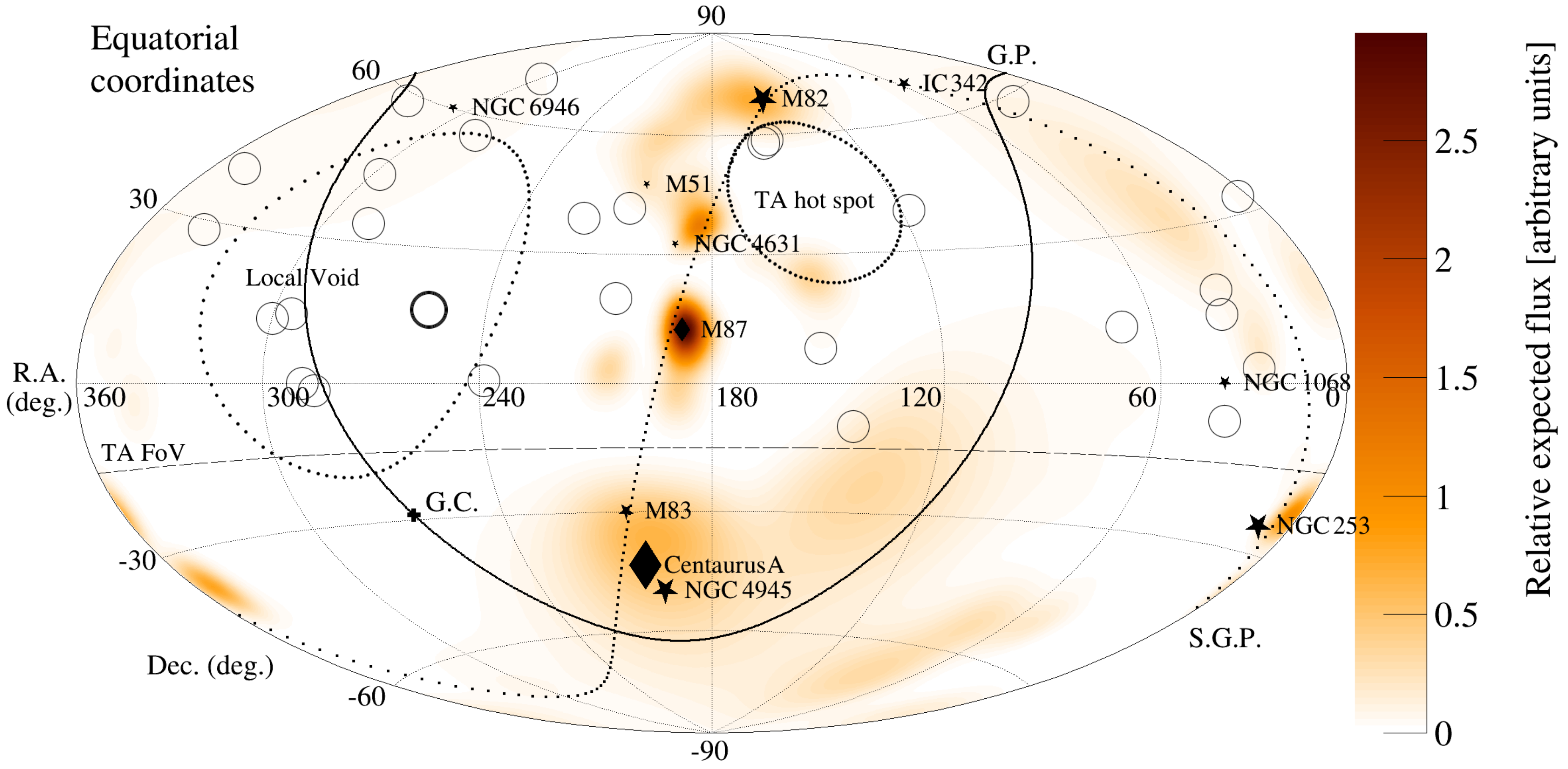}
  \caption{\textbf{Arrival directions (empty circle) of all $>$100\,EeV cosmic rays
  observed by TA SD during 13.5 years operation.} The background and other symbols are the same as Fig.~\ref{fig:highest}. No clustering around the highest energy event (thick circle) is evident.}
\label{fig:100EeV}
\end{figure}
Fig.~\ref{fig:100EeV} shows the arrival directions for the 28 TA SD events with energy $>$100\,EeV observed between 2008 May and 2021 November, using the same event selection~\cite{bib:hotspot_ta}. 
The total exposure is $1.6\times10^4$\,km$^2$\,sr\,yr. No clustering with the highest energy event is found. 
The 244\,EeV event came from a different direction than the previously reported TA hot spot, a 3.4$\sigma$ excess centered at (Right ascension (R.A.), Declination (Dec.)) = (146.7$^{\circ}$, 43.2$^{\circ}$) found for events with energies $>$57\,EeV~\cite{bib:hotspot_ta}.

Although we may have expected events with energies above 100\,EeV to be clustered, the observed arrival directions above 100\,EeV have an isotropic distribution (Fig.~\ref{fig:100EeV}). 
The lack of a nearby source for the 244\,EeV event could be due to larger magnetic deflections than predicted by the GMF models caused by a heavy primary particle or stronger magnetic fields than in the models.
Alternatively, super-GZK UHECRs could be indicating an incomplete understanding of particle physics.
If there are unknown types of primary particles that are immune to the interactions with the CMB, they could retain their energy while traveling to Earth from more distant active galaxies.
We can not distinguish between these possibilities with the observed events.

\section*{Summary and conclusions}
We detected a particle with an energy of 244\,$\pm$\,29\,(stat.) $^{+51}_{-76}$\,(syst.)\,EeV on 27 May 2021.
The arrival direction of this event does not align with any known astronomical objects thought to be a potential source of UHECRs, even after taking into account deflection by the GMF under various assumptions.
Comparison to other observed events at energies above 100\,EeV shows an isotropic distribution with no apparent clustering.

\bibliography{scibib}

\begin{thebibliography}{10}

\bibitem{Hillas:1985is}
A.~M. Hillas, {\it Ann. Rev. Astron. Astrophys.\/} {\bf 22}, 425 (1984).

\bibitem{Bhattacharjee:1999mup}
P.~Bhattacharjee, G.~Sigl, {\it Phys. Rept.\/} {\bf 327}, 109 (2000).

\bibitem{Stecker:2009hj}
F.~W. Stecker, S.~T. Scully, {\it New J. Phys.\/} {\bf 11}, 085003 (2009).

\bibitem{Berezinsky:1997hy}
V.~Berezinsky, M.~Kachelriess, A.~Vilenkin, {\it Phys. Rev. Lett.\/} {\bf 79},
  4302 (1997).

\bibitem{TelescopeArray:2018rbt}
R.~U. Abbasi, {\it et~al.\/}, {\it Astropart. Phys.\/} {\bf 110}, 8 (2019).

\bibitem{PierreAuger:2022uwd}
P.~Abreu, {\it et~al.\/}, {\it Astrophys. J.\/} {\bf 933}, 125 (2022).

\bibitem{Linsley:1963km}
J.~Linsley, {\it Phys. Rev. Lett.\/} {\bf 10}, 146 (1963).

\bibitem{Penzias:1965wn}
A.~A. Penzias, R.~W. Wilson, {\it Astrophys. J.\/} {\bf 142}, 419 (1965).

\bibitem{bib:gzk1}
K.~Greisen, {\it Phys. Rev. Lett.\/} {\bf 16}, 748 (1966).

\bibitem{bib:gzk2}
G.~Zatsepin, V.~Kuzmin, {\it Journal of Experimental and Theoretical Physics
  Letters\/} {\bf 4}, 78 (1966).

\bibitem{bib:hires_gzk}
R.~Abbasi, {\it et~al.\/}, {\it Phys. Rev. Lett.\/} {\bf 100}, 101101 (2008).

\bibitem{AbuZayyad:2012ru}
T.~Abu-Zayyad, {\it et~al.\/}, {\it Astrophys. J.\/} {\bf 768}, L1 (2013).

\bibitem{PierreAuger:2020kuy}
A.~Aab, {\it et~al.\/}, {\it Phys. Rev. Lett.\/} {\bf 125}, 121106 (2020).

\bibitem{AbuZayyad:2012kk}
T.~Abu-Zayyad, {\it et~al.\/}, {\it Nucl. Instrum. Meth.\/} {\bf A689}, 87
  (2013).

\bibitem{Tokuno:2012mi}
H.~Tokuno, {\it et~al.\/}, {\it Nucl. Instrum. Meth.\/} {\bf A676}, 54 (2012).

\bibitem{TelescopeArray:2018xyi}
R.~U. Abbasi, {\it et~al.\/}, {\it Astrophys. J.\/} {\bf 858}, 76 (2018).

\bibitem{TelescopeArray:2018bep}
R.~U. Abbasi, {\it et~al.\/}, {\it Phys. Rev. D\/} {\bf 99}, 022002 (2019).

\bibitem{Heck:1998vt}
D.~Heck, J.~Knapp, J.~N. Capdevielle, G.~Schatz, T.~Thouw, {\it
  Forschungszentrum Karlsruhe Report (FZKA)\/} {\bf 6019} (1998).

\bibitem{TheTelescopeArray:2015mgw}
R.~U. Abbasi, {\it et~al.\/}, {\it Astropart. Phys.\/} {\bf 80}, 131 (2016).

\bibitem{sm}
{\it Materials and methods are available as supplementary materials\/} .

\bibitem{bib:hotspot_ta}
R.~Abbasi, {\it et~al.\/}, {\it Astrophys.J.\/} {\bf 790}, L21 (2014).

\bibitem{LyndonEvans_2008}
L.~Evans, P.~Bryant, {\it Journal of Instrumentation\/} {\bf 3}, S08001 (2008).

\bibitem{TAgamma_ICRC2021}
O.~Kalashev, {\it et~al.\/}, {\it Proceedings of Science\/} {\bf International
  Cosmic Ray Conference 2021}, 864 (2021).

\bibitem{TAML_ICRC2021}
I.~Kharuk, {\it et~al.\/}, {\it Proceedings of Science\/} {\bf International
  Cosmic Ray Conference 2021}, 384 (2021).

\bibitem{vaisala}
 {\bf https://lightning-exporter.vaisala.com}.

\bibitem{Bird:1994uy}
D.~J. Bird, {\it et~al.\/}, {\it Astrophys. J.\/} {\bf 441}, 144 (1995).

\bibitem{Hayashida:1994hb}
N.~Hayashida, {\it et~al.\/}, {\it Phys. Rev. Lett.\/} {\bf 73}, 3491 (1994).

\bibitem{Sakaki:2001md}
N.~Sakaki, {\it et~al.\/}, {\it {Proceedings of 27th International Cosmic Ray
  Conference}\/} (2001), p. 337.

\bibitem{PierreAuger:2022axr}
P.~Abreu, {\it et~al.\/}, {\it Astrophys. J.\/} {\bf 935}, 170 (2022).

\bibitem{spectrumwg_ptep}
V.~Verzi, D.~Ivanov, Y.~Tsunesada, {\it Progress of Theoretical and
  Experimental Physics\/} {\bf 2017}, 12A103 (2017).

\bibitem{Jansson:2012pc}
R.~Jansson, G.~R. Farrar, {\it Astrophys. J.\/} {\bf 757}, 14 (2012).

\bibitem{Pshirkov:2011um}
M.~S. Pshirkov, P.~G. Tinyakov, P.~P. Kronberg, K.~J. Newton-McGee, {\it
  Astrophys. J.\/} {\bf 738}, 192 (2011).

\bibitem{AlvesBatista:2016vpy}
R.~Alves~Batista, {\it et~al.\/}, {\it Journal of Cosmology and Astroparticle
  Physics\/} {\bf 05}, 038 (2016).

\bibitem{Fermi-LAT:2019yla}
S.~Abdollahi, {\it et~al.\/}, {\it Astrophys. J. Suppl.\/} {\bf 247}, 33
  (2020).

\bibitem{Farrar:2008ex}
G.~R. Farrar, A.~Gruzinov, {\it Astrophys. J.\/} {\bf 693}, 329 (2009).

\bibitem{Sowards-Emmerd_2005}
D.~Sowards-Emmerd, R.~W. Romani, P.~F. Michelson, S.~E. Healey, P.~L. Nolan,
  {\it The Astrophysical Journal\/} {\bf 626}, 95 (2005).

\bibitem{bib:sbg_auger}
A.~Aab, {\it et~al.\/}, {\it Astrophys. J.\/} {\bf 853}, L29 (2018).

\bibitem{Tully:2007ue}
R.~B. Tully, {\it et~al.\/}, {\it Astrophys. J.\/} {\bf 676}, 184 (2008).

\bibitem{MapConstruction}
M.~Y. Kuznetsov, P.~G. Tinyakov, {\it Journal of Cosmology and Astroparticle
  Physics\/} {\bf 04}, 065 (2021).

\bibitem{Abbasi:2018tqo}
R.~U. Abbasi, {\it et~al.\/}, {\it Astrophys. J.\/} {\bf 867}, L27 (2018).

\bibitem{10.1093/mnrasl/slz030}
J.~J. Eldridge, L.~Xiao, {\it Monthly Notices of the Royal Astronomical Society
  Letters\/} {\bf 485}, L58 (2019).

\bibitem{Kalashev:2014xna}
O.~E. Kalashev, E.~Kido, {\it J. Exp. Theor. Phys.\/} {\bf 120}, 790 (2015).

\bibitem{data_release}
 {\bf https://doi.org/10.5281/zenodo.8427755}.

\end{thebibliography}

\bibliographystyle{Science}

\newpage 

\section*{Acknowledgments}

The Telescope Array experimental site became available through the cooperation of the Utah School and Institutional Trust Lands Administration (SITLA), U.S. Bureau of Land Management (BLM), and the U.S. Air Force. We appreciate the assistance of the State of Utah and Fillmore offices of the BLM in crafting the Plan of Development for the site.  Patrick A.~Shea assisted the collaboration with valuable advice and supported the collaboration’s efforts. The people and the officials of Millard County, Utah have been a source of steadfast and warm support for our work which we greatly appreciate. We are indebted to the Millard County Road Department for their efforts to maintain and clear the roads. 
We gratefully acknowledge the contribution from the technical staffs of our home institutions. 
We gratefully acknowledge allocation of computer time from the Center for High Performance Computing at the University of Utah. 
We thank to Robert~Cady for his long-standing contribution in construction and operation of the detector, and to Rosa~Mayta for her development of the event viewer tool.
T.~Fujii acknowledges insightful and productive discussions in the cosmic-ray group of Kyoto University and interdisciplinary communications in the Hakubi Center for Advanced Research, Kyoto University, and in the program for the Development of Next-generation Leading Scientists with Global Insight (L-INSIGHT). 

\subsection*{Funding}
The Telescope Array experiment is supported by the Japan Society for
the Promotion of Science (JSPS) through Grants-in-Aid for Priority Area
431,
for Specially Promoted Research
JP21000002,
for Scientific  Research (S)
JP19104006,
for Specially Promoted Research
JP15H05693,
for Scientific  Research (S)
JP15H05741, for Science Research (A) JP18H03705,
for Young Scientists (A)
JPH26707011,
and for Fostering Joint International Research (B)
JP19KK0074,
by the joint research program of the Institute for Cosmic Ray Research (ICRR), The University of Tokyo;
by the Pioneering Program of RIKEN for the Evolution of Matter in the Universe (r-EMU);
by the U.S. National Science
Foundation awards PHY-1607727, PHY-1712517, PHY-1806797, PHY-2012934, and PHY-2112904;
by the National Research Foundation of Korea
(2017K1A4A3015188, 2020R1A2C1008230, 2020R1A2C2102800);
by the Ministry of Science and Higher Education of the Russian Federation under the contract 075-15-2020-778, IISN project No. 4.4501.18; Belgian Science Policy under IUAP VII/37 (ULB); and by the Simons Foundation (00001470, NG).
The Telescope Array was partially supported by the grants of the joint research program of the Institute for Space-Earth Environmental Research, Nagoya University and Inter-University Research Program of the Institute for Cosmic Ray Research of University of Tokyo. 
Also funded by the foundations of Dr. Ezekiel R. and Edna Wattis Dumke, Willard L. Eccles, and George S. and Dolores Dor\'e Eccles.
The State of Utah supported the Telescope Array through its Economic Development Board, and the University of Utah through the Office of the Vice President for Research. 

\subsection*{Author Contributions}
T.~Fujii identified the event, performed data analyses and wrote the manuscript. R.~Higuchi assisted the back-tracking calculation, T.~Sako contributed the Monte Carlo simulations, M. Yu.~Kuznetsov calculated the relative expected flux, and I.~Kharuk performed the proton-gamma-ray classification.
J.N.~Matthews, P.~Sokolsky, G.B.~Thomson, H.~Sagawa, C.H.~Jui, S.~Ogio, Y.~Tsunesada, S.V.~Troitsky, G.I.~Rubtsov, P.G.~Tinyakov, J.H.~Kim, K. Fujita and N.~Globus discussed the results and commented on the manuscript.
All authors meet the journal’s authorship criteria for their contributions to the detector construction, deployment, long-term data-taking and maintenance, software development and review of the manuscript.

\subsection*{Competing interests}
There are no competing interests to declare. 

\subsection*{Data and materials availability}
Data of the extremely energetic cosmic ray event and codes are available at Zenodo~\cite{data_release}.

%

\newpage

\section*{Telescope Array Collaboration Authors and Affiliations}
R.U.~Abbasi$^{1}$,
M.G.~Allen$^{2}$,
R.~Arimura$^{3}$,
J.W.~Belz$^{2}$,
D.R.~Bergman$^{2}$,
S.A.~Blake$^{4}$,
B.K.~Shin$^{5}$,
I.J.~Buckland$^{2}$,
B.G.~Cheon$^{6}$,
T.~Fujii$^{7,3,8\ast}$,  
K.~Fujisue$^{9}$,
K.~Fujita$^{9}$,
M.~Fukushima$^{9}$,
G.D.~Furlich$^{2}$,
Z.R.~Gerber$^{2}$,
N.~Globus$^{10,\ddag}$,
K.~Hibino$^{11}$,
R.~Higuchi$^{10}$,
K.~Honda$^{12}$,
D.~Ikeda$^{11}$,
H.~Ito$^{10}$,
A.~Iwasaki$^{3}$,
S.~Jeong$^{13}$,
H.M.~Jeong$^{13}$,
C.H.~Jui$^{2}$,
K.~Kadota$^{14}$,
F.~Kakimoto$^{11}$,
O.E.~Kalashev$^{15}$,
K.~Kasahara$^{16}$,
K.~Kawata$^{9}$,
I.~Kharuk$^{15}$,
E.~Kido$^{10}$,
S.W.~Kim$^{13}$,
H.B.~Kim$^{6}$,
J.H.~Kim$^{2}$,
J.H.~Kim$^{17}$,
I.~Komae$^{3}$,
Y.~Kubota$^{18}$,
M.Yu.~Kuznetsov$^{15}$,
K.H.~Lee$^{13}$,
B.K.~Lubsandorzhiev$^{15}$,
J.P.~Lundquist$^{19}$,
J.N.~Matthews$^{2}$,
S.~Nagataki$^{10}$,
T.~Nakamura$^{18}$,
A.~Nakazawa$^{18}$,
T.~Nonaka$^{9}$,
S.~Ogio$^{9}$,
M.~Ono$^{20,10}$,
H.~Oshima$^{9}$,
I.H.~Park$^{13}$,
M.~Potts$^{2}$,
S.~Pshirkov$^{15}$,
J.R.~Remington$^{21}$,
D.C.~Rodriguez$^{22,2}$,
C.~Rott$^{2,13}$,
G.I.~Rubtsov$^{15}$,
D.~Ryu$^{5}$,
H.~Sagawa$^{9}$,
N.~Sakaki$^{10}$,
T.~Sako$^{9}$,
N.~Sakurai$^{23}$,
H.~Shin$^{9}$,
J.D.~Smith$^{2}$,
P.~Sokolsky$^{2}$,
B.T.~Stokes$^{2}$,
T.S.~Stroman$^{2}$,
K.~Takahashi$^{9}$,
M.~Takeda$^{9}$,
A.~Taketa$^{24}$,
Y.~Tameda$^{25}$,
S.~Thomas$^{2}$,
G.B.~Thomson$^{2}$,
P.G.~Tinyakov$^{26}$,
I.~Tkachev$^{15}$,
T.~Tomida$^{18}$,
S.V.~Troitsky$^{15}$,
Y.~Tsunesada$^{3,8}$,
S~Udo$^{11}$,
F.R.~Urban$^{27}$,
T.~Wong$^{2}$,
K.~Yamazaki$^{28}$,
Y.~Yuma$^{18}$,
Y.V.~Zhezher$^{15}$
and Z.~Zundel$^{2}$\\
\\
$^{1}$ Physics Department, Loyola University Chicago, Chicago, IL, USA \\
$^{2}$ High Energy Astrophysics Institute and Department of Physics and Astronomy, University of Utah, Salt Lake City, Utah 84112-0830, USA\\
$^{3}$ Graduate School of Science, Osaka Metropolitan University, 3-3-138 Sugimoto, Sumiyoshi, Osaka, 558-8585, Japan\\
$^{4}$ Stellar Science, Albuquerque, New Mexico, 87110, USA \\
$^{5}$ Department of Physics, Ulsan National Institute of Science and Technology, 44919, Ulsan, Korea\\
$^{6}$ Department of Physics and The Research Institute of Natural Science, Hanyang University, Seongdong-gu, Seoul, Korea\\
$^{7}$ Hakubi Center for Advanced Research and Graduate School of Science, Kyoto University, Sakyo, Kyoto, 606-8502, Japan\\
$^{8}$ Nambu Yoichiro Institute of Theoretical and Experimental Physics, Osaka Metropolitan University, 3-3-138 Sugimoto, Sumiyoshi, Osaka, 558-8585, Japan\\
$^{9}$ Institute for Cosmic Ray Research, University of Tokyo, 5-1-5 Kashiwanoha, Kashiwa-shi, Chiba, 277-8582, Japan\\
$^{10}$ Institute of Physical and Chemical Research, 2-1 Hirosawa, Wako, Saitama, 351-0198 Japan\\
$^{11}$ Faculty of Engineering, Kanagawa University, 3-27-1 Rokkakubashi, Kanagawa-ku, Yokohama 221-8686, Japan\\
$^{12}$ University of Yamanashi, Kofu, 400-8510, Japan\\
$^{13}$ Department of Physics, SungKyunKwan University, Jang-an-gu, Suwon 16419, Korea\\
$^{14}$ Department of Natural Sciences, Tokyo City University,  Setagaya-ku, Tokyo 158-8557, Japan\\
$^{15}$ Institute for Nuclear Research of the Russian Academy of Sciences, prospekt 60-letiya Oktyabrya 7a, Moscow 117312, Russia\\
$^{16}$ Shibauta Institute of Technology and Sicence, Fukasaku  307, Minuma-ku, Saitama, Japan\\
$^{17}$ Physics Division, Argonne National Laboratory, Lemont, Illinois 60439, USA\\
$^{18}$ Academic Assembly School of Science and Technology Institute of Engineering, Shinshu University, Nagano, Nagano, 380-8553, Japan\\
$^{19}$ Center for Astrophysics and Cosmology, University of Nova Gorica, Nova Gorica, Slovenia\\
$^{20}$ Institute of Astronomy and Astrophysics, Academia Sinica, Taipei 10617, Taiwan \\
$^{21}$ NASA Marshall Space Flight Center, Martin Road, Huntsville, AL 35808, USA\\
$^{22}$ Integrated Support Center for Nuclear Nonproliferation and Nuclear Security, Japan Atomic Energy Agency, Tokai-mura, Ibaraki 319-1195, Japan\\
$^{23}$ Faculty of Design Technology, 3-1-1 Nakagaito, Daito City, Osaka, Japan\\
$^{24}$ Earthquake Research Institute, University of Tokyo, Bunkyo-ku, Tokyo, 113-0032, Japan \\
$^{25}$ Department of Engineering Science, Faculty of Engineering, Osaka Electro-Communication University, Neyagawa-shi, Osaka 572-8530, Japan\\
$^{26}$ Universite Libre de Bruxelles, bvd du Triomphe CP225, Brussels, Belgium\\
$^{27}$ The Central European Institute for Cosmology and Fundamental Physics, Institute of Physics of the Czech Academy of Sciences, Na Slovance 1999/2, 182 21 Prague, Czech Republic\\
$^{28}$ College of Engineering, Chubu University, 1200 Matsumoto, Kasugai, Aichi 487-8501, Japan\\
$\ddag$ Present address: Department of Astronomy and Astrophysics, University of California, Santa Cruz, CA 95064, USA\\
$^\ast$ E-mail: toshi@omu.ac.jp \\

\end{document}